\documentclass[superscriptaddress,groupedaddress,nofootnoteinbib,12pt]{article} 
\pdfoutput=1
\usepackage{graphicx}
\usepackage{color}
\usepackage{dcolumn}   
\usepackage{bm}       
\usepackage{amssymb}  
\usepackage{amsmath}
\usepackage{sectsty}
\usepackage{latexsym}
\usepackage{float}
\usepackage{ifthen}
\usepackage{caption,subfig}
\usepackage{enumerate}
\usepackage{url}
\usepackage{caption,subfig}
\usepackage{feynmf}	
\usepackage{jcappub}
\usepackage{amsopn}
\usepackage{float}
\usepackage{pst-pdf}
\usepackage{amsfonts}
\usepackage{multirow}
\usepackage{array}
\usepackage{booktabs}
\usepackage{rotating}

\def\clap#1{\hbox to 0pt{\hss#1\hss}}

\def\({\left(}
\def\){\right)}
\def\[{\left[}
\def\]{\right]}
\def\bea{\begin{eqnarray}}
\def\eea{\end{eqnarray}}
\def\be{\begin{equation}}
\def\ee{\end{equation}}
\def\ba{\begin{eqnarray}}
\def\ea{\end{eqnarray}}
\def\beq{\begin{eqnarray}}
\def\eeq{\end{eqnarray}}

\def\clap#1{\hbox to 0pt{\hss#1\hss}}

\definecolor{forestgreen}{rgb}{0.133,0.545,0.133}
\newboolean{editorial}
\setboolean{editorial}{true}
\newcommand{\editorial}[2]{\ifthenelse{\boolean{editorial}}{\textcolor{red}{[\textsf{\textbf{{#1}}}: }\textcolor{blue}{\textsf{{#2}}}\textcolor{red}{]}}{}}

\renewcommand{\vec}[1]{\bm{\mathrm{{#1}}}}

 \def\be   {\begin{equation}}   \def\ee   {\end{equation}}

 \def\ba  {\begin{eqnarray}}   \def\ea  {\end{eqnarray}}

\renewcommand{\thesubsection}{\arabic{section}.\arabic{subsection}}

\begin{document}

\title{Generalization of the Proca Action}

\author{Lavinia Heisenberg,$^{a,b}$}
\affiliation{$^{a}$Perimeter Institute for Theoretical Physics, \\
31 Caroline St. N, Waterloo, Ontario, Canada, N2L 2Y5}
\affiliation{$^{b}$D\'epartment de Physique  Th\'eorique and Center for Astroparticle Physics,\\
Universit\'e de Gen\`eve, 24 Quai E. Ansermet, CH-1211  Gen\`eve, Switzerland}

	\emailAdd{Lavinia.Heisenberg@unige.ch}

\abstract{We consider the Lagrangian of a vector field with derivative self-interactions with a priori arbitrary coefficients. Starting with a flat space-time we show that  for a special
choice of the coefficients of the self-interactions the ghost-like pathologies disappear. For this we use the degeneracy condition of the Hessian. This constitutes the Galileon-type generalization of the Proca action with only three propagating physical degrees of freedom. The longitudinal mode of the vector field is associated to the usual Galileon interactions for a specific choice of the overall functions. In difference to a scalar Galileon theory, the generalized Proca field has more free parameters and purely intrinsic vector interactions. We then extend this analysis to a curved background. The resulting theory is the Horndeski Proca action with second order equations of motion on curved space-times. }


\maketitle


\section{Introduction}

Motivated by the work from de Rham and Gabadadze for the generalization of the Fierz-Pauli action for a massive graviton \cite{deRham:2010ik}, we investigate here the generalization of the Proca action for a massive vector field with derivative self-interactions. We will be addressing the natural question of what is the Lagrangian for a self-interacting vector field with second order equations of motion yielding three propagating physical degrees of freedom. We will call them the "vector Galileons", since they contain derivative self-interactions for the vector field and the longitudinal mode corresponds to a Galileon \cite{Nicolis:2008in}.

In the standard relativistic quantum field theory we describe particles with local covariant field operators like scalars, vectors, tensors..etc. The finite-dimensional representation of the Lorentz group dictates to us the number of propagating degrees of freedom. For a massless spin-1 field the theory needs to have the gauge symmetry in order to have the Lorentz invariance manifestly built in. The theory describes then a massless spin-1 field with two propagating degrees of freedom $h=\pm 1$. On the other hand, for a massive spin-1 field we have three propagating physical degrees of freedom (as it also becomes clear from the master formula $(2s+1)$, where $s$ represents the spin of the particle). The Proca action is the theory describing  a massive vector field, which propagates the corresponding three polarizations (two transverse plus one longitudinal) . The mass term breaks explicitly the $U(1)$ gauge invariance such that the longitudinal mode propagates as well. However, the zero component of the vector field does not propagate. So therefore it is a natural question to investigate also the existence of derivative interactions for the vector field with still only three propagating degrees of freedom and turning the temporal component manifestly non-dynamical. This is exactly what we aim in this paper: we want to find the generalization of the Proca action for a massive vector field with derivative self-interactions. 

The standard Proca action is given by
\begin{equation}
\mathcal S_{\rm Proca}  = \int \mathrm{d}^4x \left[-\frac14 F_{\mu\nu}^2 -\frac12 m^2 A^2 \right]  \label{ProcaAction} 
\end{equation}
where $F_{\mu\nu}=\partial_\mu A_\nu -\partial_\nu A_\mu$. In this theory, the temporal component of the vector field does not propagate and generates a primary constraint. The consistency condition of this primary constraint generates a secondary constraint, whose Poisson bracket with the primary constraint is proportional to the mass, so that only in the massless case it corresponds to a first class constraint generating a gauge symmetry. In the presence of the mass term it represents a second class constraint. 

Gauge invariance is just a redundancy in the description of massless particles describing the same physical state. Therefore, we can use the Stueckelberg trick to restore the gauge invariance. In the case of the standard Proca action we can restore the gauge invariance by adding an additional scalar field via $A_\mu \to A_\mu + \partial_\mu \pi$. This trick does not change the number of propagating physical degrees of freedom. We add one additional scalar degree of freedom but we restore the gauge invariance which guaranties the existence of only two physical degrees of freedom for the vector field, making it in total three physical degrees of freedom. When we now take the mass going zero limit  $m \to0$ the Lagrangian results in a theory of a massless scalar field completely decoupled from a massless vector field for a conserved source. This is the reason why there is no vDVZ discontinuity in the case of the Proca field for conserved sources. This is different for the massive graviton. There, the helicity-0 degree of freedom does not decouple and gives rise to an additional fifth force which has to be screened via the Vainshtein mechanism. In the case of the vector field, we do not need any Vainshtein mechanism since there is no observational difference between a massless and massive vector fields for conserved sources. 

In the generalized Proca action that we construct here, the longitudinal mode of the vector field has exactly the same interactions as a Galileon scalar field for a specific choice of the overall functions. The Galileon theory is an important class of infra-red modifications of general relativity . The Galileon interactions were introduced as a natural extension of the decoupling limit of the DGP model \cite{Nicolis:2008in}. It is constructed as an effective field theory for a scalar field by the restriction of the invariance under internal Galilean and shift transformations and second order equations of motion. This effective action is local and contains higher order derivatives. Nevertheless, these interactions come in a very specific way such that they only give rise to second order equations of motion. The allowed interactions for the Galileon were originally determined order by order by writing down all the possible contractions for the derivative scalar field interactions and finding the proper coefficients giving rise to second order equations of motion. However, de Rham and Tolley could construct an unified class of four dimensional effective theories starting from a higher dimensional setup and show that these effective theories reproduce successfully all the interaction terms of the Galileon in the non-relativistic limit \cite{deRham:2010eu}. In a similar way we wonder whether or not one could construct the generalized Proca action that we are proposing here from a higher dimensional set-up which we will investigate in a future work. Naively, we would think, that, starting from a higher dimensional set-up with manifestly covariant Lovelock invariants, one would only construct terms which are gauge invariant after dimensional reduction. There is only one possible non-minimal interaction which fulfills this requirement, namely the contraction of two field strength tensors with the dual Riemann tensor. Therefore, this specific interaction with gauge invariance could be easily constructed from a higher dimensional set-up. However, it would be worth to study, if the other not gauge invariant non-minimal couplings could be constructed by dimensional reduction.

The Galileon interactions present a subclass of Horndeski interactions which describe scalar-tensor interactions with at most second order equations of motion on curved backgrounds \cite{Horndeski:1974wa}. Interestingly, a subclass of Horndeski scalar-tensor interactions \cite{deRham:2011by} can also be constructed by covariantizing the decoupling limit of massive gravity \cite{deRham:2010kj}. In the literature there has been some attempts to find a theory for vector fields which is equivalent to scalar Galileons, i.e. to find the {\it vector Galileons} besides the Maxwell kinetic term with second order
equations of motion on flat space-times \cite{Deffayet:2013tca}. There, the authors were interested in derivative self-interactions for the vector field with gauge symmetry yielding only two propagating degrees of freedom. They concluded that the Maxwell kinetic term is the only allowed interaction and wrote a no-go theorem for generalized vector Galileons \footnote{A similar no-go theorem has been also studied in \cite{deRham:2013tfa} in the context of massive graviton.}. However, on curved backgrounds there only exists one term respecting the gauge symmetry, which is given by the non-minimal coupling between the field strength tensor and the double dual Riemann tensor \cite{Horndeski:1976gi, Jimenez:2013qsa}. However, if one gives up on the gauge invariance, meaning that we allow for terms which are not invariant under $A_\mu \to A_\mu + \partial_\mu\theta$, then one can indeed construct {\it vector Galileons} on flat-spacetimes or Horndeski vector interactions on curved backgrounds giving rise to three propagating physical degrees of freedom with second order equations of motion. We will illustrate this in this work.

\section{The theory of generalized Proca field}
\label{sec:dyn_analy}
Now we want to generalize the Proca action \ref{ProcaAction} to include derivative self-interactions of the vector field, but without changing the number of propagating degrees of freedom.  In order to obtain such interactions, we will analyze all the possible Lorentz invariant terms that can be built at each order and constrain the interactions to remove the ghost-instabilities. The Lagrangian for the generalized Proca vector field with derivative self-interactions is given by 
\begin{equation}\label{generalizedProfaField}
\mathcal L_{\rm gen. Proca} = -\frac14 F_{\mu\nu}^2 +\sum^5_{n=2}\alpha_n \mathcal L_n \,,
\end{equation}
where the self-interactions of the vector field are
\begin{eqnarray}\label{vecGalProcaField}
\mathcal L_2 & = &f_2(A_\mu, F_{\mu\nu}, \tilde{F}_{\mu\nu})\nonumber\\
\mathcal L_3 & = &f_3(A^2) \;\; \partial\cdot A \nonumber\\
\mathcal L_4  &=&  f _4(A^2)\;\left[(\partial\cdot A)^2-\partial_\rho A_\sigma \partial^\sigma A^\rho\right] + c_2\tilde{f} _4(A^2)F^2 \nonumber\\
\mathcal L_5  &=&f_5(A^2)\;\left[(\partial\cdot A)^3-3(\partial\cdot A)\partial_\rho A_\sigma \partial^\sigma A^\rho +2\partial_\rho A_\sigma \partial^\gamma A^\rho\partial^\sigma A_\gamma  \right]  \nonumber\\
&& 
+d_2\tilde{f}_5(A^2)\tilde{F}^{\mu\alpha}\tilde{F}_{\alpha}^\nu\partial_\mu A_\nu  \nonumber\\
\mathcal L_6  &=&e_2f_6(A^2) \tilde{F}^{\alpha\beta}\tilde{F}^{\mu\nu}\partial_\alpha A_\mu \partial_\beta A_\nu \,,
\end{eqnarray}
with $\partial\cdot A=\partial_\mu A^\mu$ and where the functions $f_{2,3,4,5}$ are arbitrary functions. Let us first emphasize the dependences of these functions. First of all, they all can depend on $A^2=A_\mu A^\mu$. Nevertheless, the function $f_2$ is special in the sense that it is the only function which is not multiplied by any term with derivatives acting on the vector field. Therefore, this function $f_2$ can also have dependence on all the possible terms which have $U(1)$ symmetry like $F^2=F^{\mu\nu}F_{\mu\nu}$, $F\tilde{F}=F^{\mu\nu}\tilde{F}_{\mu\nu}$..etc (where $\tilde{F}$ is the dual of F). Furthermore, the function $f_2$ can depend on terms which does not contain any time derivative applying on the temporal component  $A_0$ of the vector field like for instance $A_\mu A_\nu F^{\mu \rho}F_\rho^\nu$. This is not true for the remaining functions $f_{3,4,5}$. Thus, we have
\begin{equation}
f_{2} =  f_{2}(A_\mu, F_{\mu\nu}, \tilde{F}_{\mu\nu})  \hspace{0.5cm} \text{and} \hspace{0.5cm} f_{3,4,5}=f_{3,4,5}(A^2)
\end{equation}
For instance the function $f_{2}$ can naturally depend on terms like $A^2 F^2$, $A^2 F^4$, $A^4 F^4$, $A_\mu A_\nu F^{\mu \rho}F_\rho^\nu$ ...etc while the remaining functions $f_{3,4,5}$ can only depend on $A^2$ since these functions are multiplied by terms which contain derivatives acting on the vector field. These functions do not change the number of propagating physical degrees of freedom since they do not contain any dynamics for the temporal component of the vector field. We will comment more on that in section \ref{sec_special_case}. The second Lagrangian $\mathcal L_2$ naturally contains the mass $\frac12 m^2 A^2$ and potential terms $V(A^2)$ for the vector field in the function $f_2$. 
In the next section we will illustrate order by order why these interactions give rise to only three propagating degrees of freedom and illustrate the absence of ghost instabilities.
 Note also the appearance of the three free parameters $c_2$, $d_2$ and $e_2$. It means that the vector Galileons contain more free parameters then the usual scalar Galileon theory. 
 We would like to mention that the sixth order Lagrangian was first omitted in a previous version of this work, since by then it was imposed that the longitudinal mode should not have any trivial total derivative. Getting rid of this condition permitted the construction of the sixth order Lagrangian in \cite{Allys:2015sht}, which was confirmed in \cite{Jimenez:2016isa}. An interesting possibility for an infinite series of derivative self-interactions was also discussed in \cite{Allys:2015sht}, which unfortunately trivializes in four dimensions \cite{Jimenez:2016isa}.
 
 The interactions can be also expressed in terms of the Levi-Civita tensors 
\begin{eqnarray}\label{epsilon1}
\mathcal L_2&=&-\frac{f_2(A_\mu, F_{\mu\nu}, \tilde{F}_{\mu\nu})}{24}\mathcal{E}^{\mu\nu\alpha\beta}\mathcal{E}_{\mu\nu\alpha\beta} =f_2(A_\mu, F_{\mu\nu}, \tilde{F}_{\mu\nu}) \nonumber\\
\mathcal L_3&=&-\frac{f_3(A^2)}{6}\mathcal{E}^{\mu\nu\alpha\beta}\mathcal{E}^{\rho}_{\;\;\;\nu\alpha\beta} \partial_\mu A_\rho=f_3(A^2)\;\; \partial\cdot A \nonumber\\
\mathcal{L}_4&=&-\frac{1}{2}\mathcal{E}^{\mu\nu\rho\sigma}\mathcal{E}^{\alpha\beta}_{\;\;\;\;\rho\sigma}(f_4(A^2)\partial_\mu A_\alpha\partial_\nu A_\beta+c_2\tilde{f}_4(A^2)\partial_\mu A_\nu\partial_\alpha A_\beta)\nonumber\\
&=&f _4(A^2)\; \left[(\partial\cdot A)^2-\partial_\rho A_\sigma \partial^\sigma A^\rho\right]+c_2 \tilde{f} _4(A^2)F_{\rho\sigma}^2 \nonumber\\
\mathcal{L}_5&=&-\mathcal{E}^{\mu\nu\rho\sigma}\mathcal{E}^{\alpha\beta\delta}_{\;\;\;\;\;\;\sigma}(f_5(A^2)\partial_\mu A_\alpha\partial_\nu A_\beta \partial_\rho A_\delta +d_2\tilde{f}_5(A^2)\partial_\mu A_\nu\partial_\rho A_\alpha \partial_\beta A_\delta) \nonumber\\
& =&f_5(A^2)\left[(\partial\cdot A)^3-3(\partial\cdot A)\partial_\rho A_\sigma \partial^\sigma A^\rho 
+2\partial_\rho A_\sigma \partial^\gamma A^\rho\partial^\sigma A_\gamma \right] \nonumber\\
&+&d_2\tilde{f}_5(A^2)\tilde{F}^{\alpha\mu}\tilde{F}^\beta_{\;\;\mu}\partial_\alpha A_\beta \nonumber\\
\mathcal{L}_6&=&-\mathcal{E}^{\mu\nu\rho\sigma}\mathcal{E}^{\alpha\beta\delta\kappa}(f_6(A^2)\partial_\mu A_\alpha\partial_\nu A_\beta \partial_\rho A_\delta \partial_\sigma A_\kappa +e_2\tilde{f}_6(A^2)\partial_\mu A_\nu\partial_\alpha A_\beta \partial_\rho A_\delta  \partial_\sigma A_\kappa)\nonumber\\
&=&f_6(A^2)\left[3\partial^\beta A^\alpha(\partial_\alpha A_\beta \partial_\mu A_\nu \partial^\nu A^\mu-2\partial_\alpha A_\nu \partial^\mu A_\beta \partial^\nu A_\mu)+8(\partial\cdot A)\partial_\beta A_\nu \partial^\mu A^\beta \partial^\nu A_\mu \right. \nonumber\\
&&\left.-6 (\partial\cdot A)^2\partial_\mu A_\nu \partial^\nu A_\mu+(\partial\cdot A)^4 \right] \nonumber\\
&+&e_2\tilde{f}_6(A^2)\Big[ (\partial\cdot A)(2\partial^\gamma A^\beta(\partial_\beta A_\delta \partial^\delta A_\gamma-\partial_\delta A_\gamma \partial^\delta A_\beta)+(\partial\cdot A)(\partial_\delta A_\gamma - \partial_\gamma A_\delta)\partial^\delta A^\gamma)) \nonumber\\
&+&\partial^\beta A^\alpha(-2\partial_\alpha A_\delta \partial^\gamma A_\beta \partial^\delta A_\beta+\partial_\gamma A_\delta(2\partial^\gamma A_\alpha \partial^\delta A_\beta+(\partial_\alpha A_\beta -\partial_\beta A_\alpha)\partial^\delta A^\gamma))\Big]\nonumber\\
&=&e_2\tilde{f}_6(A) \tilde{F}^{\alpha\beta}\tilde{F}^{\mu\nu}\partial_\alpha A_\mu \partial_\beta A_\nu \,.
\end{eqnarray}
In \cite{Allys:2015sht,Jimenez:2016isa} the first contraction in $\mathcal{L}_6$ was then neglected since it corresponds to a total derivative (see also \cite{Heisenberg:2018vsk}). Motivated by the interesting phenomenology and implications of the generalized Proca theories \cite{Tasinato:2014eka,DeFelice:2016yws,DeFelice:2016uil,Chagoya:2017fyl,Heisenberg:2017hwb}, this term was resurrected in \cite{ErrastiDiez:2019ttn}. However, as it has been explicitly shown in \cite{Jimenez:2019hpl} this term is indeed just a total derivative in four-dimensional flat space-time and hence can be ignored.
The Lagrangians $\mathcal L_{2,3,4,5,6}$ in (\ref{vecGalProcaField}) propagate only three degrees of freedom. Higher order interactions beyond the six order Lagrangian are trivial in four dimensions, hence the series stops here. Expressed in terms of the Levi-Civita tensors this means that we run out of the indices.\\ 
When we wrote the derivative self-interactions in terms of the Levi-Civita tensors, the indices of the potential interactions were always contracted with each other. Without loss of generality consider for example the special choice for the functions $f_{2,3,4,5}=(A^2)$ then in this case, we could either consider contractions in the functions as $A_\mu A^\mu$ or contract the indices of these two vector fields with the Levi-Civita tensor as well. One might wonder, if it yields different interactions once the indices of the term $(A^2)$ for example are contracted with the Levi-Civita tensors as well. 
\begin{eqnarray}
\mathcal L_2^{al}&=&-\frac16\mathcal{E}^{\mu\nu\alpha\beta}\mathcal{E}^{\rho}_{\;\;\;\nu\alpha\beta}A_\mu A_{\rho} =(A^2) \nonumber\\
\mathcal L_3^{al}&=&-\frac12\mathcal{E}^{\mu\nu\alpha\beta}\mathcal{E}^{\rho\sigma}_{\;\;\;\alpha\beta}A_\mu A_{\rho} \partial_\nu A_\sigma=(A^2)(\partial\cdot A)-A^\mu A^\nu \partial_\nu A_\mu \nonumber\\
\mathcal L_4^{al}&=&-\mathcal{E}^{\mu\nu\alpha\beta}\mathcal{E}^{\rho\sigma\delta}_{\;\;\;\;\;\beta}A_\mu A_{\rho} \partial_\nu A_\sigma \partial_\alpha A_\delta \nonumber\\
&=&(A^2)\left[(\partial\cdot A)^2-\partial_\rho A_\sigma \partial^\sigma A^\rho\right] -2A^\mu A^\nu \partial_\nu A_\mu(\partial\cdot A)+2A^\mu A^\nu \partial_\nu A_\rho \partial^\rho A_\mu \nonumber \\
\mathcal L_5^{al}&=&\mathcal{E}^{\mu\nu\alpha\beta}\mathcal{E}^{\rho\sigma\delta\gamma}A_\mu A_{\rho} \partial_\nu A_\sigma \partial_\alpha A_\delta\partial_\beta A_\gamma \nonumber\\
&=&(A^2)\left[-(\partial\cdot A)^3+3(\partial\cdot A)\partial_\rho A_\sigma \partial^\sigma A^\rho-2\partial_\rho A_\sigma \partial^\gamma A^\rho\partial^\sigma A_\gamma\right]\nonumber\\
&& +3A^\mu A^\nu \partial_\nu A_\mu(\partial\cdot A)^2-6A^\mu A^\nu \partial_\nu A_\rho \partial^\rho A_\mu(\partial\cdot A)+6A^\mu A^\nu \partial_\nu A_\rho \partial^\rho A_\gamma \partial^\gamma A_\mu \nonumber\\
&&-3A^\mu A^\nu \partial_\nu A_\mu \partial_\rho A_\sigma \partial^\sigma A^\rho
\end{eqnarray}
But on closer inspection one can see that they give rise to exactly the same interactions once integrations by part are performed. This would just correspond to considering disformal transformations of the metric $\eta_{\mu\nu}\to \eta_{\mu\nu}+A_\mu A_\nu$. For instance if we take the cubic interaction $\mathcal{L}_3$ in \eqref{epsilon1} and perform a disformal transformation $\eta_{\mu\nu}\rightarrow \eta_{\mu\nu}+\tilde{f}_3(X)A_\mu A_\nu$, then this will give rise to $\tilde{\mathcal{L}}_3=\tilde{f}_3(X)A^\mu A^\nu (\partial_\mu A_\nu)$. We do not consider these interactions as genuinely new interactions since they are related to the previous interactions by means of disformal transformations. In the following we will start with the general interactions order by order with arbitrary coefficients and demonstrate that imposing the absence of the unphsical degree of freedom gives rise to the above Lagrangian with only three propagating physical degrees of freedom. This will make extensive use of the degeneracy condition of the Hessian and the presence of second class contraints.


\section{The propagation of three degrees of freedom}
The simplest modification of the Proca action \ref{ProcaAction} is of course promoting the mass term to an arbitrary function $f_2$ which contains amongst others the mass term and the potential interactions for the vector field $f_2 \supset V(A^2)$, since this trivially does not modify the number of degrees of freedom. As we already emphasized, this function can also contain gauge invariant interactions which are invariant under the $U(1)$ transformations and terms which do not contain any dynamics for the temporal component of the vector field, i.e. terms of the form $f_2 \supset F^2 + FF^*+A^2F^2+A^2FF^*+A_\mu A_\nu F^{\rho\mu}F_{\rho}^\nu + \cdots$. The independent contractions are $X=-A_\mu A^\mu/2$, $F=-F_{\mu\nu}F^{\mu\nu}/4$ and $Y=A^\mu A^\nu F_{\mu}{}^{\alpha}F_{\nu\alpha}$ (see also \cite{Fleury:2014qfa}) and hence we can rewrite the function as $\mathcal L_2  =f_2(X,F,Y)$.

The first term that we can have to the next order in the vector field is simply 
\begin{equation}
\mathcal L_3  = f_3(X)\;\; \partial\cdot A \label{vecGalL3}
\end{equation}
with $f_3$ being an arbitrary function of the vector field norm $X=-A_\mu A^\mu/2$. It is a trivial observation that in (\ref{vecGalL3}) the temporal component of the vector field $A_0$ does not propagate, even if we include the Maxwell kinetic term,  and it acts as a lagrange multiplier. The easiest way to see it is by computing the corresponding Hessian, which vanishes trivially $H_{\mathcal L_3}^{\mu\nu}=0$. Also notice that the presence of the function $f_3$ is crucial since if it was simply a constant, that term would be a total divergence and, thus, with no contribution to the field equations.

To next order, the independent interaction terms that we can have are given by
\begin{equation}
\mathcal L_4  =  f_4 \; \left[c_1(\partial\cdot A)^2+c_2\partial_\rho A_\sigma \partial^\rho A^\sigma+c_3\partial_\rho A_\sigma \partial^\sigma A^\rho\right]  \label{vecGalL4gen}
\end{equation}
with a priori free parameters $c_1$, $c_2$ and $c_3$ and $f_4$ an arbitrary function depending on $f_4(A^2)$. Now,  we need to fix the parameters such that only  three physical degrees of freedom propagate, i.e., such that we still have a second class constraint. In order to eliminate one propagating degree of freedom, we need a constraint equation, which is guaranteed if the determinant of the Hessian matrix vanishes. The Hessian matrix for (\ref{vecGalL4gen}) is given by
\begin{eqnarray}
H_{\mathcal L_4}^{\mu\nu}=\frac{\partial^2\mathcal{L}_4}{\partial \dot A_\mu \partial \dot A_\nu} =f_4(X)
\begin{pmatrix}
2(c_1+c_2+c_3)&0&0&0 \\
0&-2c_2 &0&0 \\
0&0&-2c_2 &0 \\
0&0&0&-2c_2
 \end{pmatrix}
\end{eqnarray}
For a vanishing determinant of the Hessian matrix we have two possibilities. First possibility corresponds to choosing $c_2=0$. In this case the Hessian matrix contains three vanishing eigenvalues corresponding to three constraints. Therefore, if we choose $c_2=0$, only the zero component of the vector field propagate while the other three degrees of freedom do not propagate. This is not what we are looking for, therefore we disregard this choice. The other possibility for a vanishing determinant of the Hessian matrix corresponds to $c_1+c_2+c_3=0$. Without loss of generality we can set $c_1=1$ and therefore $c_3=-(1+c_2)$. In this case the Hessian matrix only contains one vanishing eingenvalue and hence only one propagating constraint. This case corresponds to three propagating degrees of freedom with the Lagrangian at this order given by:
\begin{equation}
\mathcal L_4  =  f _4\; \left[(\partial\cdot A)^2+c_2\partial_\rho A_\sigma \partial^\rho A^\sigma-(1+c_2)\partial_\rho A_\sigma \partial^\sigma A^\rho\right]  \label{vecGalL4}
\end{equation}
Note that we can write these interactions also as 
\begin{equation}
\mathcal L_4  = f _4 \; \left[(\partial\cdot A)^2-\partial_\rho A_\sigma \partial^\sigma A^\rho+c_2F_{\rho\sigma}^2\right]
\end{equation}
where it becomes immediate that the last term proportional to $c_2$ even with an independent separate function $\tilde{f}_4(X)$ can then simply be absorbed into $f_2\supset F^2$ since the function $f_2$ depends in general on the gauge invariant quantities which do not contain any dynamics for the $A_0$ degree of freedom and therefore will not change the number of propagating degrees of freedom. One can either include the interaction $c_2\tilde{f}_4(X)F_{\rho\sigma}^2$ into the function $f_2$ or leave it at the order of $\mathcal L_4$, but not both at the same to avoid redundancy. In terms of the Levi-Civita tensors we have the following ways of contracting the indices
\begin{eqnarray}
\mathcal{L}_4&=&-\frac{1}{2}\mathcal{E}^{\mu\nu\rho\sigma}\mathcal{E}^{\alpha\beta}_{\;\;\;\;\rho\sigma}(f_4(X)\partial_\mu A_\alpha\partial_\nu A_\beta+c_2\tilde{f}_4(X)\partial_\mu A_\nu\partial_\alpha A_\beta)\nonumber\\
&=&f _4\; \left[(\partial\cdot A)^2-\partial_\rho A_\sigma \partial^\sigma A^\rho\right]+c_2\tilde{f} _4(\partial_\rho A_\sigma \partial^\rho A^\sigma-\partial_\rho A_\sigma \partial^\sigma A^\rho) \,.
\end{eqnarray}
The vanishing of the determinant of the Hessian matrix guaranties the existence of a constraint. To find the expression for the constraint, we have to compute the conjugate momentum $\Pi^\mu_{\mathcal L_4}=\frac{\partial \mathcal L_4}{\partial\dot A_\mu}$. The zero component of the conjugate momentum is given by
\begin{equation}
\Pi^0_{\mathcal L_4}=-2f_4\; \vec{\nabla}\vec{A}\,.
\end{equation}
As one can see, the zero component of the conjugate momentum does not contain any time derivative  yielding the constraint equation 
\begin{equation}
\mathcal C_1=\Pi^0_{\mathcal L_4}+2f_4\; \vec{\nabla}\vec{A}. 
\end{equation}
This constraint equation will generate a secondary constraint given by
\begin{equation}
\{ H, \mathcal C_1\}=\frac{\partial H}{\partial A_\mu}\frac{\partial \mathcal C_1}{\partial \Pi^\mu}-\frac{\partial H}{\partial \Pi^\mu}\frac{\partial \mathcal C_1}{\partial A_\mu}
\end{equation}
or equivalently one can obtain the secondary constraint by calculating the time derivative of the conjugate momentum $\dot \Pi^\mu$ and use the Hamiltonian equations $\frac{\partial H}{\partial A_\mu}=-\dot \Pi^\mu$ and $\frac{\partial H}{\partial \Pi^\mu}=\dot A_\mu$. We have checked explicitly the existence of the secondary constraint  and therefore the Lagrangian $\mathcal L_4$ possesses only three propagating degrees of freedom.
\\
For the next order interactions we write down all the possible contractions between the derivative self-interactions which gives:
\begin{eqnarray}
\mathcal L_5  =  f_5 &&\left[d_1(\partial\cdot A)^3-3d_2(\partial\cdot A)\partial_\rho A_\sigma \partial^\rho A^\sigma-3d_3(\partial\cdot A)\partial_\rho A_\sigma \partial^\sigma A^\rho \right. \nonumber\\
&&\left. +2d_4\partial_\rho A_\sigma \partial^\gamma A^\rho\partial^\sigma A_\gamma+2d_5\partial_\rho A_\sigma \partial^\gamma A^\rho\partial_\gamma A^\sigma\right] \label{vecGalL5gen}
\end{eqnarray}
with a priori the arbitrary parameters $d_1$, $d_2$, $d_3$, $d_4$ and $d_5$ and function $f_5$ depending only on $A^2$. In this quintic Lagrangian (\ref{vecGalL5gen}) the additional possible term $\partial_\sigma A_\rho \partial^\gamma A^\rho\partial_\gamma A^\sigma$ is actually equal to $\partial_\rho A_\sigma \partial^\gamma A^\rho\partial_\gamma A^\sigma$ since $ \partial^\gamma A^\rho\partial_\gamma A^\sigma$ is symmetric under the exchange of $\rho$ and $\sigma$. The Hessian matrix for this quintic Lagrangian is giving by
\begin{eqnarray}
H_{\mathcal L_5}^{00}&=&-6(d_1-d_2-d_3)(\vec{\nabla}\vec{A})+6(d_1-3d_2-3d_3+2(d_4+d_5))\dot A_t \nonumber\\
H_{\mathcal L_5}^{0i}&=&H_{\mathcal L_5}^{i0}=(6d_3-2(3d_4+d_5))A_{t,i}+2(3d_2-2d_4)A_{i,t} \nonumber\\
H_{\mathcal L_5}^{11}&=&-6d_2 A_\alpha^2(A_{z,z}+A_{y,y})-2(3d_2-2d_4)(A_{x,x}-A_{t,t}) \nonumber\\
H_{\mathcal L_5}^{12}&=&H_{\mathcal L_5}^{21}=2d_5(A_{x,y}+A_{y,x}) \nonumber\\
H_{\mathcal L_5}^{13}&=&H_{\mathcal L_5}^{31}=2d_5(A_{x,z}+A_{z,x}) \nonumber\\
H_{\mathcal L_5}^{22}&=&2(-3d_2A_{z,z}+(-3d_2+2d_5)A_{y,y}-3d_2A_{x,x}+(3d_2-2d_5)A_{t,t}) \nonumber\\
H_{\mathcal L_5}^{23}&=&H_{\mathcal L_5}^{32}=2d_5(A_{y,z}+A_{z,y}) \nonumber\\
H_{\mathcal L_5}^{33}&=&(-6d_2+4d_5)A_{z,z}-6d_2(A_{y,y}+A_{x,x})+2(3d_2-2d_5)A_{t,t}
\end{eqnarray}
In order to have only three propagating degrees of freedom the parameters need to fulfill the following conditions
\begin{eqnarray}
&&d_1-d_2-d_3=0, \;\;\;\;\;\;\;\;\;\;\;\;d_1-3d_2-3d_3+2(d_4+d_5)=0, \nonumber \\
&&3d_3-3d_4-d_5=0,\;\;\;\;\;\; \;\;\;3d_2-2d_5=0
\end{eqnarray}
which are fulfilled by choosing (again without loss of generality we can choose $d_1=1$)
\begin{eqnarray}
d_3=1-d_2, \;\;\;\;\;\; \;\; d_4=1-\frac{3d_2}{2},  \;\;\;\;\;\;\;d_5=\frac{3d_2}{2}
\end{eqnarray}
Hence, the quintic Lagrangian with only three propagating physical degrees of freedom is given by
\begin{eqnarray}
\mathcal L_5  =  f_5 &&\left[(\partial\cdot A)^3-3d_2(\partial\cdot A)\partial_\rho A_\sigma \partial^\rho A^\sigma-3(1-d_2)(\partial\cdot A)\partial_\rho A_\sigma \partial^\sigma A^\rho \right. \nonumber\\
&&\left. +2\left(1-\frac{3d_2}{2}\right)\partial_\rho A_\sigma \partial^\gamma A^\rho\partial^\sigma A_\gamma+2\left(\frac{3d_2}{2}\right)\partial_\rho A_\sigma \partial^\gamma A^\rho\partial_\gamma A^\sigma\right] \label{vecGalL5}
\end{eqnarray}
Analogously, we can write these interactions also as
\begin{eqnarray}
\mathcal L_5  &=&  f_5 \left[(\partial\cdot A)^3-3(\partial\cdot A)\partial_\rho A_\sigma \partial^\sigma A^\rho+2\partial_\rho A_\sigma \partial^\gamma A^\rho\partial^\sigma A_\gamma \right. \nonumber\\
&& \left. -\frac{3d_2}{2}(\partial\cdot A)F_{\rho\sigma}^2+3d_2\partial_\sigma A_\gamma F_\rho^{\;\;\sigma}F^{\rho\gamma}\right].
\end{eqnarray}
The Hessian matrix with this chosen parameters then becomes 
\begin{eqnarray}
H_{\mathcal L_5}^{\mu\nu} =f_5(A^2)
\begin{pmatrix}
0&0&0&0 \\
0&-6d_2 (A_{z,z}+A_{y,y})&3d_2(A_{x,y}+A_{y,x})&3d_2(A_{x,z}+A_{z,x}) \\
0&3d_2 (A_{x,y}+A_{y,x})&-6d_2(A_{z,z}+A_{x,x})&3d_2(A_{y,z}+A_{z,y}) \\
0&3d_2 (A_{x,z}+A_{z,x})&3d_2 (A_{y,z}+A_{z,y})&-6d_2 (A_{y,y}+A_{x,x})
 \end{pmatrix}
\end{eqnarray}
with a vanishing determinant $\det{(H^{\mu\nu}_{\mathcal L_5})}=0$. As required, the Hessian matrix only contains one vanishing eingenvalue and hence only one propagating constraint which is again given by the corresponding zero component of the conjugate momentum $\Pi^\mu_{\mathcal L_5}=\frac{\partial \mathcal L_5}{\partial\dot A_\mu}$ 
\begin{eqnarray}
\Pi^0_{\mathcal L_5}&=&-3f_5(A^2) \left(d_2(A_{x,z}^2+A_{y,z}^2+A_{x,y}^2)-2A_{z,z}A_{y,z}-2(-1+d_2)A_{y,z}A_{z,y}+d_2A_{z,y}^2 +d_2A_{z,x}^2 \right. \nonumber \\ 
&& \left. -2(A_{z,z}+A_{y,y})A_{x,x} +2A_{x,y}A_{y,x}-2d_2A_{x,y}A_{y,x} + d_2A_{y,x}^2-2(-1+d_2)A_{x,z}A_{z,x} \right)\,.
\end{eqnarray}
As you can see, there is no time derivatives appearing in the expression of the zero component of the conjugate momentum, representing the constraint equation. Associated to this constraint, there will be a secondary constraint guarenting the propagation of the constraint equation and removing the unphysical degree of freedom.
Expressed with the Levi-Civita tensors, the distinctive nature of the interactions become transparent
\begin{eqnarray}
\mathcal{L}_5&=&-\mathcal{E}^{\mu\nu\rho\sigma}\mathcal{E}^{\alpha\beta\delta}_{\;\;\;\;\;\;\sigma}(f_5(X)\partial_\mu A_\alpha\partial_\nu A_\beta \partial_\rho A_\delta +d_2\tilde{f}_5(X)\partial_\mu A_\nu\partial_\rho A_\alpha \partial_\beta A_\delta) \nonumber\\
& =&f_5(X)\left[(\partial\cdot A)^3-3(\partial\cdot A)\partial_\rho A_\sigma \partial^\sigma A^\rho 
+2\partial_\rho A_\sigma \partial^\gamma A^\rho\partial^\sigma A_\gamma \right] \nonumber\\
&+&d_2\tilde{f}_5(X)\tilde{F}^{\alpha\mu}\tilde{F}^\beta_{\;\;\mu}\partial_\alpha A_\beta\,.
\end{eqnarray}
Note that the longitudinal part of the vector field belongs to the Galileon scalar interactions. Imposing the condition that the longitudinal mode should not have any trivially vanishing interactions would make that the series for the longitudinal mode would stop here. However, since the vector field contains two transverse modes besides the longitudinal mode, one can construct the interactions forth order in derivates of the vector field, that would give rise to a non-trivial mixing between the transverse and the longitudinal modes \cite{Allys:2015sht,Jimenez:2016isa} as it was the case for the $d_2\tilde{f}_5(X)\tilde{F}^{\alpha\mu}\tilde{F}^\beta_{\;\;\mu}\partial_\alpha A_\beta$ interaction in $\mathcal{L}_5$. The sixth order Lagrangian contains
\begin{align}
\mathcal{L}_6=&-\mathcal{E}^{\mu\nu\rho\sigma}\mathcal{E}^{\alpha\beta\delta\kappa}(f_6(X)\partial_\mu A_\alpha\partial_\nu A_\beta \partial_\rho A_\delta \partial_\sigma A_\kappa +e_2\tilde{f}_6(X)\partial_\mu A_\nu\partial_\alpha A_\beta \partial_\rho A_\delta  \partial_\sigma A_\kappa)\nonumber\\
&=e_2\tilde{f}_6(X) \tilde{F}^{\alpha\beta}\tilde{F}^{\mu\nu}\partial_\alpha A_\mu \partial_\beta A_\nu\,.
\end{align}
The first contraction corresponds to just a total derivative \cite{Heisenberg:2018vsk,Jimenez:2019hpl} and can be neglected.


\section{Galileon case of the functions $f_{2,3,4,5}=A^2$}\label{sec_special_case}
In this section we will pay attention to the special case where the arbitrary functions are chosen to be $f_{2,3,4,5}=A^2$ in the Lagrangians $L_{2,3,4,5}$. Considering only those Lagrangians in \ref{vecGalProcaField} simplifies to
\begin{eqnarray}\label{vecGalProcaFieldspecial}
\mathcal L_2 & = &A^2\nonumber\\
\mathcal L_3 & = &A^2(\partial\cdot A) \nonumber\\
\mathcal L_4  &=& A^2\left[(\partial\cdot A)^2+c_2\partial_\rho A_\sigma \partial^\rho A^\sigma-(1+c_2)\partial_\rho A_\sigma \partial^\sigma A^\rho\right]   \nonumber\\
\mathcal L_5  &=& A^2\left[(\partial\cdot A)^3-3d_2(\partial\cdot A)\partial_\rho A_\sigma \partial^\rho A^\sigma-3(1-d_2)(\partial\cdot A)\partial_\rho A_\sigma \partial^\sigma A^\rho \right. \nonumber\\
&&\left. +2\left(1-\frac{3d_2}{2}\right)\partial_\rho A_\sigma \partial^\gamma A^\rho\partial^\sigma A_\gamma +2\left(\frac{3d_2}{2}\right)\partial_\rho A_\sigma \partial^\gamma A^\rho\partial_\gamma A^\sigma\right] \,.
\end{eqnarray}
We can restore the $U(1)$ gauge symmetry using the Stueckelberg trick by adding an additional scalar field via $A_\mu \to A_\mu +\partial_\mu \pi$. To zeroth order in $A_\mu$ we extract out only the longitudinal mode of the vector field and recover exactly the Galileon interactions
\begin{eqnarray}
\mathcal L_2 & = & (\partial\pi)^2\nonumber\\
\mathcal L_3 & = &(\partial\pi)^2\Box\pi\nonumber\\
\mathcal L_4 & = & (\partial\pi)^2\left[(\Box\pi)^2-(\partial_\mu\partial_\nu\pi)^2\right]\nonumber\\
\mathcal L_5 & = & (\partial\pi)^2\left[(\Box\pi)^3-3\Box\pi(\partial_\mu\partial_\nu\pi)^2+2(\partial_\mu\partial_\nu\pi)^3\right]\,.
\end{eqnarray}
Note that after introducing the gauge symmetry the dependence of the free parameters $c_2$ and $d_2$ disappears in the purely longitudinal sector as expected. Similarly, to first order in $A_\mu$ we obtain the following scalar-vector interactions
\begin{eqnarray}
\mathcal L_2 & = & 2A^\mu \partial_\mu \pi \nonumber\\
\mathcal L_3 & = &(\partial\pi)^2(\partial\cdot A) +2\Box\pi\partial_\mu \pi A^\mu \nonumber\\
\mathcal L_4 & = & 2(\partial\pi)^2\Box\pi(\partial\cdot A)+2(\Box\pi)^2\partial_\mu \pi A^\mu-2(\partial\pi)^2\partial_\mu\partial_\nu \pi \partial^\nu A^\mu-2(\partial_\mu \partial_\nu \pi)^2\partial_\rho \pi A^\rho \nonumber\\
\mathcal L_5 & = & 6(\partial\pi)^2\partial_\sigma\partial_\beta\pi \partial_\alpha \partial^\sigma \pi \partial^\beta A^\alpha+2A^\alpha \partial_\alpha \pi (\Box\pi)^3+3(\partial\cdot A)(\partial\pi)^2(\Box\pi)^2-6\partial_\alpha \partial_\beta \pi \partial^\beta A^\alpha (\partial\pi)^2\Box\pi \nonumber\\
&+&4A^\alpha \partial_\alpha\pi \partial^\rho \partial^\beta \pi \partial_\sigma \partial_\beta \pi \partial^\sigma \partial_\rho \pi -3\Big(2A^\alpha \partial_\alpha\pi \Box\pi+(\partial\cdot A)(\partial\pi)^2\Big)(\partial_\rho\partial_\sigma\pi)^2
\end{eqnarray}
This is another way of observing that the interactions we found for the vector field indeed only propagate three degrees of freedom, since when we plug in the longitudinal mode, we obtain the Galileon interaction with at most second order equations of motion. The terms for the vector field $\partial_\rho A_\sigma \partial^\rho A^\sigma$ and $\partial_\rho A_\sigma \partial^\sigma A^\rho$ are not the same, but when we replace $A_\mu=\partial_\mu\pi$, they are since the derivatives acting on the scalar field commute $\partial_\mu\partial_\nu\pi=\partial_\nu\partial_\mu\pi$ on flat space-time. This has a huge concequence: \emph{the interactions for the vector field have more free parameters than the Galileon interactions}. It means that if we had started with the Galileon interactions and performed the replacement $\partial_\mu \pi \to A_\mu$ we would have been missing some of the interactions which also yield three propagating degrees of freedom. The vector interactions have three more free parameters (namely what we called $c_2$, $d_2$ and $e_2$ in \eqref{vecGalProcaField}). In fact, an alternative way of finding our generalized Proca action is by restoring the $U(1)$ gauge invariance and imposing that the Stueckelberg field propagates only one degree of freedom, i.e., it satisfies second order field equations. One must be careful though, since in addition to the pure Stueckelberg sector, it is also necessarily to analyse the terms mixing the Stueckelberg field and the vector field. For $\mathcal L_4$, no additional constraints arise from the mixing terms, since we obtain terms of the general form $K^{\mu\nu}(A_\mu)\partial_\mu\pi\partial_\nu\pi$, which automatically leads to second order contributions for $\pi$. However,  for $\mathcal L_5$ we obtain terms like $K^{\alpha\beta\gamma\delta}(A_\mu)\partial_\alpha\partial_\beta\pi\partial_\gamma\partial_\delta\pi$ so we need to impose the tensor $K^{\alpha\beta\gamma\delta}(A_\mu)$ to have the correct structure. This Stueckelberg analysis in the decoupling limit and the conditions for the $K$ tensors of the allowed mixing between the Stueckelberg field and the transverse fields have been analyzed in detail in \cite{Jimenez:2016isa}.

It is also worth to emphasize one more time that the arbitrary functions $f_{2,3,4,5}$ appearing in our generalized Proca action have been chosen to be $A^2$ in this section to be able to relate them to the Galileon interactions. In the Stueckelberg language, this is so in order to guarantee the second order nature of the field equations with respect to $\pi$. There are however additional contributions upon which the functions might depend without altering the number of degrees of freedom. Such terms are those for which the Stueckelberg field give a trivial contribution, i.e., those which are $U(1)$ gauge invariant. Therefore, the function $f_2$ could actually depend also on the combinations $F^2$ or $FF^*$. It can naturally also depend on any possible contraction between $A_\mu$ and $F_{\mu\nu}$ as well, in a way like for example $A_\mu A_\nu F^{\mu\alpha}F_\alpha^{\;\; \nu}$..etc. From the vector field perspective, these terms do not contain time derivatives of $A_0$, so that it will not spoil the existence of the constraints. Indeed, if you look at the interactions in $\mathcal L_4$ which are proportional to the parameter $c_2$ then you trivially recognize that these terms are just $c_2 F^2$. Since the function $f_{2}$ also depends on $F^2$ then the term for instance in $\mathcal L_4$ could be absorbed into $f_2(X,F,Y)$. One must be cautious however, since arbitrary functions of such invariants typically give rise to violations of the hyperbolicity of the field equations and hence to superluminal propagation, which we do not discuss in this work.

The equations of motion for the Lagrangian of the derivative self-interacting vector field (\ref{vecGalProcaFieldspecial}) on top of the Maxwell kinetic term are given by
\begin{eqnarray}
\mathcal E_{2} & = & 2A_\mu\nonumber\\
\mathcal E_3 & = &2A_\mu(\partial\cdot A) -2A^\nu \partial_\mu A_\nu \nonumber\\
\mathcal E_4 & = & 2\Big(A_\mu \left[(\partial\cdot A)^2-(1+c_2)\partial_\rho A_\sigma \partial^\sigma A^\rho+c_2\partial_\rho A_\sigma \partial^\sigma A^\rho\right]
+c_2 A^2(-\Box A_\mu+\partial_\nu\partial_\mu A^\nu)\nonumber\\
&&-2c_2 A^\rho \partial_\nu A_\rho \partial^\nu A_\mu-2(\partial\cdot A) A^\rho \partial_\mu A_\rho +2(1+c_2)A^\rho \partial_\nu A_\rho \partial_\mu A^\nu \Big) \nonumber\\
\mathcal E_5 & = &2A_\mu\left[(\partial\cdot A)^3+3(-1+d_2)(\partial\cdot A)\partial_\rho A_\sigma \partial^\sigma A^\rho-3d_2(\partial\cdot A)\partial_\rho A_\sigma \partial^\rho A^\sigma+(2-3d_2)\partial_\rho A_\sigma \partial^\gamma A^\rho\partial^\sigma A_\gamma \right. \nonumber\\
&& \left. +3d_2\partial_\rho A^\sigma \partial^\rho A^\gamma\partial_\sigma A_\gamma\right]
-3A^\rho\Big( -d_2(4\partial_\nu A_\rho \partial^\nu A_\mu (\partial\cdot A)-2(\partial_\nu A_\mu \partial^\nu A_\sigma +\partial^\nu A_\mu \partial_\sigma A_\nu)\partial^\sigma A_\rho  \nonumber\\
&&+A_\rho(\partial^\nu A_\mu(\partial_\sigma\partial_\nu A^\sigma -\Box A_\nu)+2(\partial\cdot A)(\Box A_\mu - \partial_\sigma\partial_\mu A^\sigma)+(\partial_\nu\partial_\mu A_\sigma-2\partial_\sigma\partial_\nu A_\mu+\partial_\sigma\partial_\mu A_\nu)\partial^\sigma A^\nu))   \nonumber\\
&&+2((\partial\cdot A)^2+((-1+d_2)\partial_\nu A_\sigma -d_2\partial_\sigma A_\nu)\partial^\sigma A^\nu)\partial_\mu A_\rho +(4(-1+d_2)\partial_\nu A_\rho (\partial\cdot A) \nonumber\\
&&+d_2A_\rho(-\partial_\sigma\partial_\nu A^\sigma+\Box A_\nu) +2((2-3d_2)\partial_\nu A_\sigma + d_2 \partial_\sigma A_\nu)\partial^\sigma A_\rho)\partial_\mu A^\nu\Big)
\end{eqnarray}
Note also that the equations of motion for the vector field does reproduce the equations of motion of the Galileon field if we take the divergence of it and replace $A_\mu=\partial_\mu \pi$.

\section{Curved space-times}
In the flat space-time the derivatives applied on the vector field were simply partial derivatives which commute. When we consider a general non-flat background the derivatives become covariant derivatives and therefore we have to add non-minimal couplings to the graviton in order to maintain second order equations of motion and healthy propagating degrees of freedom.
When we generalize the derivative self-interactions in \ref{generalizedProfaField} on a curved space-time, the Lagrangian for the generalized Proca field becomes 
\begin{equation}\label{generalizedProfaField}
\mathcal L^{\rm curved}_{\rm gen. Proca} = -\frac14 F_{\mu\nu}^2 +\sum^5_{n=2}\beta_n \mathcal L_n
\end{equation}
where now the self-interactions are encoded in the following Lagrangians
\begin{eqnarray}\label{vecGalcurv}
\mathcal L_2 & = & G_2(A_\mu,F_{\mu\nu},\tilde{F}_{\mu\nu}) \nonumber\\
\mathcal L_3 & = &G_3(X)\nabla_\mu A^\mu \nonumber\\
\mathcal L_4 & = & G_{4}(X)R+G_{4,X} \left[(\nabla_\mu A^\mu)^2-\nabla_\rho A_\sigma \nabla^\sigma A^\rho\right] \nonumber\\
\mathcal L_5 & = & G_5(X)G_{\mu\nu}\nabla^\mu A^\nu-\frac{1}{6}G_{5,X} \Big[
(\nabla\cdot A)^3 \nonumber\\
&+&2\nabla_\rho A_\sigma \nabla^\gamma A^\rho \nabla^\sigma A_\gamma -3(\nabla\cdot A)\nabla_\rho A_\sigma \nabla^\sigma A^\rho \Big] \nonumber \\
&-&g_5(X) \tilde{F}^{\alpha\mu}\tilde{F}^\beta_{\;\;\mu}\nabla_\alpha A_\beta  \nonumber \\
\mathcal L_6 & = & G_6(X)\mathcal{L}^{\mu\nu\alpha\beta}\nabla_\mu A_\nu \nabla_\alpha A_\beta 
+\frac{G_{6,X}}{2} \tilde{F}^{\alpha\beta}\tilde{F}^{\mu\nu}\nabla_\alpha A_\mu \nabla_\beta A_\nu
\end{eqnarray}
with $\nabla$ denoting the covariant derivative. The fifth order interaction $\tilde{F}^{\alpha\mu}\tilde{F}^\beta_{\;\;\mu}\nabla_\alpha A_\beta$ does not need any non-minimal coupling on curved backgrounds in order to compensate higher order equations of motion \cite{Allys:2015sht,Jimenez:2016isa}, unless the sixth order Lagrangian $\tilde{F}^{\alpha\beta}\tilde{F}^{\mu\nu}\nabla_\alpha A_\mu \nabla_\beta A_\nu$ \cite{Jimenez:2016isa}, which requires the relative tuning with the non-minimal coupling between the field strength tensors and the dual Riemann tensor considered in \cite{Jimenez:2013qsa}. These interactions give rise to the standard scalar Horndeski interactions for the longitudinal mode of the vector field. Terms like $G^{\mu\nu}A_\mu A_\nu$, which does not contain any dynamics for the temporal component of the vector field, are already contained in the above interactions after integrations by parts. All these interactions give only rise to  three propagating degrees of freedom in curved background.

\section{Summary and discussion}
In this paper we have constructed the generalized Proca action for a vector field with derivative self-interactions with only three propagating degrees of freedom. We started our analysis with the case of a flat Minkowksi spacetime. We successfully showed that for appropriate choices of the coefficients of the derivative self-interactions that generalize the Proca action, one can construct a consistent and local
theory of massive vector field without the presence of ghost-like instabilities. The resulting theory is simple and constitutes five Lagrangians for the self-interactions of the vector field. We were able to show that the constrained coefficients yield the necessary propagating constraint in order to remove the unphysical degree of freedom. These are the vector Galileons with three propagating degrees of freedom. At each order the Lagrangian has an overall function which depends on $A^2$ and the function of the quadratic Lagrangian can also depend on all the possible terms invariant under $U(1)$ symmetry like for instance $F^2$ and $FF^*$..etc. Similarly this function can also depend on any contractions between the vector field and the field strength tensor $A_\mu A_\nu F^{\mu\rho}F_\rho^\nu$ which does not contain any time derivative applied on the temporal component of the vector field. The dependence of the function $f_2$ on the gauge invariant terms or terms in which the zero component of the vector field does not have any dynamics, do not alter the number of propagating degrees of freedom. We have also shown, that these interactions have more free parameters than the corresponding scalar Galileon interactions. We then generalized our results to the case of curved space-time and obtained the corresponding Horndeski vector interactions.



\acknowledgments

We would like to thank Claudia de Rham for useful discussions. Specially, we would like to express our infinite gratitude to Jose Beltran Jimenez for his collaboration on this subject. This work is supported by the Swiss National Science Foundation. \\
\\
{\it You are the sun, so radiant and warm. The dawn cannot compete against you, the clouds cannot cover you. I am the moon trying to reflect your rays. If I can succeed this even slightly, so I'm happy}.


\bibliographystyle{JHEPmodplain}
\bibliography{references}

\end{document}